\documentclass[12pt]{iopart}
\usepackage{graphicx}
\usepackage[all]{xy}
\begin{document}
\title{Theory of enhanced second-harmonic generation \\ by the quadrupole-dipole hybrid exciton}
\author{Oleksiy Roslyak}
\address{Physics Department, The City College, CUNY\\
Convent Ave. at 138 St, New York, N.Y. 10031,
USA}
\ead{avroslyak@gmail.com}
\author{Joseph L. Birman}
\address{Physics Department, The City College, CUNY\\
Convent Ave. at 138 St, New York, N.Y. 10031,
USA}
\begin{abstract}
We report calculated substantial enhancement of the second harmonic generation (SHG) in cuprous oxide crystals resonantly hybridized with an appropriate organic material (DCM2:CA:PS 'solid-state solvent'). The quadrupole origin of the inorganic part of the quadrupole-dipole hybrid provides inversion symmetry breaking and the organic part contributes to the oscillator strength of the hybrid. We show that the enhancement of the SHG, compared to bulk cuprous oxide crystal, is proportional to the ratio of the DCM2 dipole moment and the effective dipole moment of the quadrupole transitions in the cuprous oxide. It is also inversely proportional to the line-width of the hybrid and bulk excitons. The enhancement may be regulated by adjusting the organic blend (mutual concentration of the DCM2 and CA part of the solvent) and pumping conditions(varying the angle of incidence in case of optical pumping or populating the minimum of the lower branch of the hybrid in case of electrical pumping).
\end{abstract}
\maketitle
%-------------------------------------------------------------------------------------------------------------------------------
\section{Introduction}
Considerable attention has been paid to the relatively strong optical second-harmonic generation (SHG) in thin films ($D_{4h}$ symmetry) and bulk ($O_{h}$ symmetry) of cuprous oxide crystals. Which was first addressed in the pioneering work of Shen \cite{SHEN:1996}. This effect is attributed to the electric-quadrupole $\hbar \omega_{1S}= 2.05 \; eV$ exciton effect. The quadrupole exciton has very small oscillator strength but it possess rather narrow line-width $\hbar\gamma_{1S}$. So the effect is well pronounced when the exciting laser energy is close to one $\hbar \omega_{1S}-\hbar\omega \ll \hbar\gamma_{1S}$ or two photon resonance $\hbar \omega_{1S}-2\hbar\omega \ll \hbar\gamma_{1S}$. In the dipole approximation this effect disappears \cite{ATANASOV:1994}.
%and can not be attributed to the surface effect.
\par
We propose to amplify the SHG characteristic of the $1S$ quadrupole Wannier exciton (WE) in cuprous oxide by making a hybrid with an organic Frenkel exciton (FE) (See next section for more details). The idea of resonant enhancement of some non-linear properties generic to semiconductor dipole-allowed Wannier-Mott (WE) excitons was presented in pioneering work of \cite{AGRANOVICH:1998} for the layered organic-inorganic heterostructures. It was also developed for quantum wires and dots embedded into organic shell \cite{ENGELMAN:1998},\cite{GAO:2004} or attached to dendrimer structure \cite{HUONG:2000},\cite{HUONG:2003}.
\par
In our previous work \cite{ROSLYAK:2007} we demonstrated considerable enhancement of another non-linear effect in cuprous oxide, photo-thermal bi-stability \cite{DASBACH:2004}. We demonstrated a considerable enhancement in the hysteresis-like region size (from $\mu eV$ for bulk cuprous oxide to $meV$ for the hybrid). The enhancement was attributed to the large oscillator strength of the hybrid exciton inherited from the organic part and still rather narrow line-width of the same order as the coupling. Analogous enhancement can be expected for the SHG, which is the subject of this paper.
\par
In Section 2 we propose a pump-prob experiment to reveal the SHG enhancement due to the resonant dynamical hybridization and briefly discuss relevant quadrupole hybrid exciton properties. In the next Section 3 we address the question how this resonant\footnote{The resonance occurs between the FE and WE} enhancement depends on such parameters of the system as oscillator strengths and damping of the FE and WE constituting the hybrid. Using a classical model of nonlinear coupled oscillators, we demonstrate that while the big ratio of the hybrid oscillator strength suggests many orders of the enhancement magnitude it is actually somehow reduced by the rather small coupling parameter and density of the FE.
\par
Because the FE is dynamically brought into resonance with the WE there is an important hybridization time $\tau_h$ parameter. Hence, in the Section 4, we develop more sophisticated quantum mechanical model to address the dynamics of the hybrid SHG. Namely we show that the signal enhancement drastically depends on the either the one probes the system before or  after the hybridization occurred.
%-------------------------------------------------------------------------------------------------------------------------------
\section{Proposed experimental set-up for the SHG}
\begin{figure}[htbp]
\centering
\includegraphics[width=8.6cm]{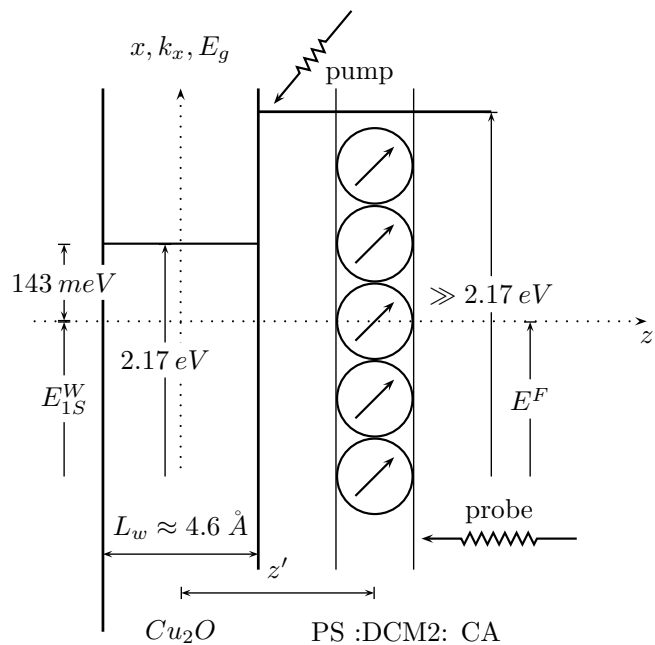}
	\caption{Schematic representation and the energy offset of a possible experimental set-up to observe the enhanced SHG by the quadrupole-dipole exciton. Here the inorganic $Cu_2O$ quantum well provides the $1S$ quadrupole WE. The DCM2 part of the organic 'solid state solute' provides dipole allowed FE (set of small arrows); the PS host prevents wave function overlapping between organic and inorganic excitons; CA under proper concentration allows tuning of the excitons into the resonance.}
	\label{FIG:1}
\end{figure}

In this work we adopt the concept of a layered organic-inorganic heterostructure. The inorganic component of the hybrid is a thin layer of $Cu_2O$ (quantum well, latter in the text referred to as QW) grown upon a film of the organic composite (See Fig.\ref{FIG:1}). Due to the small radius of both the WE and FE exciton part of the hybrid one can neglect the effect of confinement. In this case one can not tune the two types of excitons in resonance by adjusting the confinement ($L_w>a^W_B \approx$ to the cuprous oxide unit cell $a=4.6 \; \AA$). The QW confinement just assures the WE propagate along the interface and is the subjected to the electric field gradient of the FE propagating along the adjacent chain of the DCM2 molecules.
% FIGURE CONFIGURATION BEGIN
% FIGURE CONFIGURATION END
\par
To provide resonance between WE in cuprous oxide and FE in the organic, we proposed utilization of 'solid state solvation' (SSS) of the DCM2 \footnote{[2-methyl-6-2-(2,3,6,7-tetrahydro-1H, 5H - benzo[i,j] - quinolizin - 9 - yl) - ethenyl] - 4H - pyran - 4 - ylidene] propane dinitritle.} molecules in transparent polystyrene (PS) host doped with camorphic anhydride (CA)\cite{BULOVIC:1999}. The SSS is a type of solvatochromism manifesting itself as some change in the spectral position of the absorption/luminescence band due to change in the polarity of the medium. The F{\"o}rster dipole-dipole non-resonant interaction between DCM2 and CA modifies the energy structure of the involved molecules.
\par
During the 'slow' phase ($\tau_s \approx 3.3 \: ns$) the energy of the FE \footnote{in our case we define the FE as DCM2 excitation} experiences a red shift linear with the CA concentration due to non-resonant dipole-dipole interaction with the CA molecules. Not that our model capitalizes on the fact that DCM2 molecules form a 2D layer rather than been diluted in the PS:CA solvent which is the case of currently manufactured optical light emitting devices (OLED). This allows us to neglect rather complicated problem of the inhomogeneous broadening of the FE energy by utilizing a \emph{mean field} approximation \footnote{Indeed, in our simplified model the DCM2 molecules are not randomly situated but rather form a uniform (homogeneous) thin layer near the interface. Also the experimentally observed FE energy relaxation shows no significant energy fluctuations. This experiments are performed at MIT by the Dr. Bulovic. Although this results are not officially published yet, but reported in the MIT proceedings.}. For the mean field approximation the red spectral shift of the FE energy in resonance with the quadrupole WE can be accomplished with $ \rho_{CA} \approx 22 \% $ CA concentration.
\par
To avoid complicated problems of time dependent hybridization and stay within the analytical model framework, we assume that the FE and WE are in exact resonance once the DCM2 energy is in close proximity to the WE energy i.e.  $\hbar\omega_{DCM2}-\hbar \omega_{1S} \leq \Gamma_{k}$. We introduced the quadrupole-dipole coupling parameter $\Gamma_k \leq 4 \ \mu eV$ \cite{ROSLYAK:2006} (See also Appendix (\ref{EQ:1_1})). This resonant coupling gives rise to the upper and lower branches of the quadrupole-dipole hybrid (QDH) dispersion \footnote{See the eigenvalues of the linearized system (\ref{EQ:2_2}) or the Hamiltonian (\ref{EQ:3_2})}: $	\hbar \omega_{u,l}=\hbar \omega_{1S}\pm \Gamma_{\mathbf{k}}$. To populate both of the branches one needs a second pumping photon tuned into resonance with the $1S$ transition.
\par
The radiation field interacts through both dipole and quadrupole part of the hybrid. The dipole interaction can be utilized to produce linear response signal due to the pumping \cite{ROSLYAK:2007}. By using the non-linear response to the prob signal, the second harmonic can be generated through the quadrupole part of the hybrid. Different SHG regimes can be achieved by changing the timing between pumping and prob signals (See section \ref{quantum} for more details).
\par
According to the selection rules for the quadrupole-dipole hybrid the pumping signal, running along the organic-inorganic interface of the heterostructure, induces the linear polarization in the $z$ direction \cite{ROSLYAK:2006}. The prob signal induces the second order non-linear response in the cuprous oxide. Which is perpendicular to the interface, and defined by the second order polarization along the $x$ direction (See Fig.\ref{FIG:1}). The net polarization is given by a second rank tensor through the following expression:
\begin{equation}
\label{EQ:1_3}
P_z^{\left(1\right)}+P_{l=\hat{j}\times\hat{x}}^{\left(2\right)}=\chi^{(1)}_{i,z} E_i + i \chi^{(2)}_{l,i,j,x} k_x E_i E_j
\end{equation}
Here $E_i,E_j$ are the electric field of the pumping and prob lasers correspondingly. The $x$ component of the prob signal wave vector is taken to be close to zero to avoid possible interference in momentum conservation. For the sake of simplicity we are going to omit, $x$ and $l$ indexes of the tensor keeping in mind that the wave vector of the pump signal has only $x$ component and the SH signal is perpendicular to it and the prob signal polarization: $i \chi^{(2)}_{l,i,j,x} k_x = \chi^{(2)}_{i,j}$.
\par
In this paper we develop both classical and quantum mechanical models, which can be used to find a specific form of the hybrid second order nonlinear susceptibility. In section \ref{anharmonic} we demonstrate that the second order non linearity (generic to the cuprous oxide and introduced through a small parameter $\lambda$) is enhanced due to the resonant quadrupole-dipole hybridization with the organic (See (\ref{EQ:2_7})). In section \ref{quantum} we develop quantum theory of the enhanced SHG. It allows investigation of different regimes of the process defined by the time ordering between the prob pulse and the time when the FE and WE energies are close enough to form the hybrid. We generalize the concept of the double-sided Feynman diagrams \cite{MUKAMEL:1995} to include non radiative processes of the energy exchange between DCM2 and CA as well as resonant QDH between DCM2 and cuprous oxide.
%-----------------------------------------------------------------------------------------------------------------------------
\section{\label{anharmonic} Anharmonic coupled oscillators model}
As a first step, we will use the simplest classical model neglecting the non-local effects of the linear $\chi^{(1)}_{i,z}$ and non-linear susceptibility $\chi^{(2)}_{l,i,j,x}$ to describe the hybrid SHG. Namely, we adopt an extension of the anharmonic oscillator model \cite{BLOEMBERGEN:1965,MUKAMEL:1995} generalized for the case of resonant coupling between two distinct sets of oscillators. This simplified picture only covers the case when the pumping field is polarized along $\hat{z}\parallel [001]$ axis ($E_i=E_z$) and we prob the hybrid system ($\omega_{1S}=\omega_F$) with a signal perpendicular to the interface and polarized along $\hat{y}\parallel [010]$ direction ($E_j=E_y$).
\par
We consider the WE in cuprous oxide as an assembly of the oscillators with the oscillator strength per unit cell given by $f_{xz,k} \propto k_x$ (See for example reference \cite{MOSKALENKO:2002}). The second set of the oscillators with the oscillator strength given by $f^F$ corresponds to the FE in the organic.
\par
Treating the wave vector $k$ as just another parameter\footnote{In the text we are going to omit index $k$ unless we put an emphasis on it}, the polarization $P^W,P^F$ due to WE and FE can be written in terms of the effective electron-hole displacements $X,Y$ as:
\begin{eqnarray}
\label{EQ:2_1}
P^W=\frac{N}{a_B^W S} f_{xz,k} e X \\
P^F=\frac{\rho_{DCM2} N}{a_B^F S} f^F e Y
\end{eqnarray}
Here $S$ is the area of the interface and $a_B^W,a_B^F$ are the WE and FE Bohr radius, $e$ is the electron charge. The surface density of the WE and FE excitons are $N f_{xz,k} / \left({a_B S}\right)$ and $\rho_{DCM2} N f^F / \left({a_B S}\right)$ correspondingly and $N$ is the total number of the oscillators. Here we also took into account the low density ($\rho_{DCM2} = 0.05 \%$) of the DCM2 molecules in the organic to avoid the aggregation effect \cite{MADIGAN:2003}.
\par
In the time frame of the hybridization $\tau_s-\tau_h<t<\tau_s$, the WE and FE energies are at perfect resonance. Hence, the system of equations governing the oscillators dynamics can be written in the form:
\begin{eqnarray}
\label{EQ:2_2}
 \nonumber
 \ddot X + \omega _{1S}^2 X + \gamma \dot X - \frac{2\omega _{1S} \Gamma _k} {\hbar } Y - \omega _{1S}^2 \lambda X^2 = 0 \\
 \nonumber
 \ddot Y + \omega _{1S}^2 Y + \gamma \dot Y - \frac {2\omega _{1S} \Gamma _k }{\hbar }X = \frac{e}{m}E_i e^{i \omega t}\\
\end{eqnarray}
The nonlinear factor $\omega _{1S}^2 \lambda$ appears due to the prob signal. It is defined such that $\lambda$ has dimensions of reciprocal length and is considered to be small in a sense that it is much less than the reciprocal of the maximum displacement of FE ($Y$) and WE ($X$) oscillator. The exact value of $\lambda$ can be obtained either from an experiment or from the microscopic quantum theory (See next section for more details).
\par
The terms proportional to $\gamma$ describe the QDH damping. The terms proportional to $2\omega _{1S} \Gamma _k / \hbar $ describe the quadrupole-dipole coupling. Hence, the eigenvalues of the linearized system of equations (\ref{EQ:2_2}) give both branches of the QDH.
\par
The system is driven dominantly by the light-dipole interaction in the organic and the quadrupole-light interaction is neglected ($m$ is the electron mass).
\par
Using standard perturbation theory with respect to the small parameter $\lambda$ in zero order (neglecting the quadratic term) and combining equations (\ref{EQ:1_3},\ref{EQ:2_1},\ref{EQ:2_2}) one has the linear response of the hybrid and bulk cuprous oxide given by the following expressions:
\begin{eqnarray}
\label{EQ:2_3}
\nonumber
\chi^{\left( 1 \right)}_{Hy} \left( \omega  \right) = \frac{\rho_{DCM2} N}{a_B^F S}\frac{f^F e^2/m \left({\omega _{1S}^2  - \omega ^2  + i\omega \gamma }\right)}{{\left( {\omega _{1S}^2  - \omega ^2  + i\omega \gamma } \right)^2  - \left({2\omega _{1S} \Gamma _k/\hbar }\right)^2 }}\\
\nonumber
\chi ^{\left( 1 \right)}_{Cu2O} \left( \omega  \right) = \frac{N}{a_B^W S}\frac{{f_{xz,k} e^2 /m}}{{\omega _{1S}^2  - \omega ^2  + i\gamma }}
\end{eqnarray}
Including the nonlinear term as a source for the SHG to first order in the perturbation parameter, there is a displacement at $2 \omega$.
The SHG response is given by a solution of the following system:
\begin{eqnarray*}
\label{EQ:2_6}
 \ddot X + \omega _{1S}^2 X + \gamma \dot X - \frac{2\omega _{1S} \Gamma _k} {\hbar } Y - \omega _{1S}^2 \lambda X^2_{\lambda=0} = 0 \\
 \ddot Y + \omega _{1S}^2 Y + \gamma \dot Y - \frac {2\omega _{1S} \Gamma _k }{\hbar }X = 0
\end{eqnarray*}
Using the definitions (\ref{EQ:1_3}) and (\ref{EQ:2_1}) one gets the following non-linear second order response function for the hybrid and bulk cuprous oxide correspondingly:
\begin{eqnarray}
\label{EQ:2_7}
\nonumber
\chi ^{\left( 2 \right)}_{Hy} \left( 2\omega; \omega,\omega  \right) =
\frac{\rho_{DCM2} N}{a_B^F S}
 \frac{ f^F e^3/m^2 \omega^2_{1S} \lambda \left({2\omega _{1S} \Gamma _k/\hbar }\right)}{\left({\left( {\omega _{1S}^2  - \left( {2\omega } \right)^2  + i2\omega \gamma } \right)^2- \left({2\omega _{1S} \Gamma _k/\hbar }\right)^2}\right)^2} \times \\
\nonumber
\times \frac {\left( {\omega _{1S}^2  - \omega ^2  + i\omega \gamma } \right)^2}{\left( {\omega _{1S}^2  - \omega ^2  + i\omega \gamma } \right)^2  - \left({2\omega _{1S} \Gamma _k /\hbar }\right)^2}\\
%\nonumber
\chi ^{\left( 2 \right)}_{Cu2O} \left( {2\omega ;\omega ,\omega } \right) = \frac{N}{a_B^W S}\frac{{f_{xz,k} e^3 /m^2 \omega _{1S}^2 \lambda}}{{\left( {\omega _{1S}^2  - \left( {2\omega } \right)^2  + i2\gamma } \right)\left( {\omega _{1S}^2  - \omega ^2  + i\gamma } \right)^2 }}
\end{eqnarray}
Straightforward comparison of the expressions above evinces the resonant rise of the second order nonlinearity owing to hybridization. There are several competing factors involved. The enhancement by means of big oscillator strength ratio $f^F/f_{xz,k}$ is reduced by rather small coupling parameter $\Gamma_K$ and small DCM2 density $\rho_{DCM2}$ (see more numerical details in Section 5).
%-----------------------------------------------------------------------------------------------------------------------------
\section{\label{quantum} Quantum theory of SHG due to the QDH}
Although the system of non-linear susceptibilities (\ref{EQ:2_7}) in principle solves the problem of SHG due to the hybrid it does not clarify the origin of the nonlinearity $\lambda$. Also, such an important parameter as the hybridization time $\tau_h$ is left out of the classical description. Hence, in this section we propose a unified quantum theory of the hybrid SHG.
\par
The linear response of the hybrid is due to dipole transitions from the ground $\left| g \right\rangle$ state\footnote{when no excitations are present in the system} to the FE $\left| F \right\rangle$ in the organic and due to quadrupole transitions to the WE $\left| {1S} \right\rangle$ in the cuprous oxide.
The non-linearities are the result of some intermediate inter-band transitions in the cuprous oxide \cite{MUKAMEL:1995}.
\par
In cuprous oxide the nearest state in energy to the quadrupole ortho-exciton $\hbar \omega_{1S} \; \left({\Gamma^+_5}\right)$ is the $\hbar \omega_{2P} \; \left({\Gamma^-_4}\right)$ dipole allowed excitonic band $\left| 2P \right\rangle$, $E_g > \hbar \omega_{2P}> \hbar \omega_F > \hbar \omega_{1S}$. Hence it plays the main role in formation of the non-linear response and can be excited by the properly tuned prob signal. We neglect all the rest of inter-band and intra-band\footnote{due to small radius of the quadrupole WE} transitions. Therefore, the states above form a complete basis for the SHG problem:
\begin{equation}
\label{basis}
\left| g \right\rangle, \ \left| 1S \right\rangle, \ \left| F \right\rangle, \ \left| 2P \right\rangle
\end{equation}
Inversion symmetry of the DCM2 is also broken by the CA induced local field and the interface effect. Therefore, unlike in classical model, the contribution from the organic to the SHG has to be consider as well. But due to the smallness of the symmetry breaking local field it contributes a little to the SHG enhancement.
\par
Using the basis above let us introduce creation operators for the FE and the $1S$ and the $2P$ WE exciton $b^\dag = \left| F \right\rangle \left\langle g \right|$, $B^\dag_{1S} = \left| 1S \right\rangle \left\langle g \right|$, $B^\dag_{2P} = \left| 2P \right\rangle \left\langle g \right|$ respectively. The commutation algebra of the operators is presented in the Appendix (\ref{algebra}).
\par
The net polarization of the sample is defined as \cite{MUKAMEL:1995}:
\begin{eqnarray}
\label{EQ:3_1}
P = \mu^i_{1S,k} \left({B^\dag_{1S} + B_{1S}}\right) + \mu^i_{2P} \left({B^\dag_{2P} + B_{2P}}\right) \\
\nonumber
+ \mu^i_F \left({b^\dag + b}\right) + \mu^j_{1S,2P} \left({B^\dag_{1S} B_{2P} + B_{1S} B^\dag_{2P}}\right)
\end{eqnarray}
Here $\mu^i_{1S,k} = \hat{i}\cdot\hat{z} \; k_x Q_{x,z} = 3\cdot10^{-5} (k_x/k_{0,x}) \; D$ is an effective dipole moment \cite{MOSKALENKO:2002,ROSLYAK:2006} due to the quadrupole transitions associated with the oscillator strength; $k_0$ is the resonant wave vector for bulk cuprous oxide( Appendix (\ref{EQ:1_0})). The dipole moment of the transitions from $\left|{1S}\right\rangle$ to $\left\langle {2P}\right|$ is defined by \cite{ARTONI:2002,ELLIOT:1961}:
\begin{equation*}
\left({\mu^j_{1S,2P}}\right)^2 = \frac{N e^2 \hbar^2 f_{2P} }{S a_B^W 2 m^\star E_{2P}} \left({\hat{j}\times\hat{x}}\right)^2 = 6 \cdot 10^{-3} \; D^2
\end{equation*}
Finally, the DCM2 dipole moment of the transition from $\left|g\right\rangle$ to $\left\langle {F}\right|$ per unit area of the interface is given by \cite{MADIGAN:2004}:
\begin{equation*}
\left({\mu^i_F}\right)^2 = \frac{\rho_{DCM2} N e^2 \hbar^2 f^F }{S a_B^F 2 m^{*} \hbar \omega_{1S}} = 0.2 \; D^2
\end{equation*}
\par
Using equation (\ref{EQ:3_1}) and the rotating wave approximation for the resonant wave vector $k$, the hybrid Hamiltonian can be written as:
\begin{eqnarray}
\label{EQ:3_2}
H = \hbar \omega_F b^\dag  b  + \hbar \omega_{1S} B^\dag_{1S}  B_{1S} + E_{2P} B^\dag_{2P}  B_{2P}  + \Gamma _k \left( {B^\dag_{1S}  b + B_{1S} b^\dag  } \right)+ \\
\nonumber
+ \mu^i_F \left( { b^\dag E^\dag_i + b E_{i}} \right)  + \mu^i_{1S,k} \left( { B^\dag_{1S} E^\dag_i + B_{1S} E_{i}} \right)  + \mu^i_{2P} \left( { B^\dag_{2P} E^\dag_i + B_{2P} E_{i}} \right) +\\
\nonumber
+ \mu^j_{1S,2P} \left( {B^\dag_{1S}  B_{2P} E^\dag_j + B_{1S} B^\dag_{2P} E_{j}} \right)
\end{eqnarray}
The linear response from \emph{both} branches of the hybrid may be observed by pumping the hybrid with two signals $E_i||\hat{z} \propto e^{i \omega t}$. The first photon $\hbar \omega = E_{DCM2}$ excites DCM2 molecules. During the time period $\tau_s-\tau_h$ the system relaxes to the FE exciton energy close to $\hbar \omega_{1S}$ thus providing resonance between WE and FE. Then the second pumping photon $\hbar \omega = \hbar \omega_{1S}$ enters and excites quadrupole WE so that both QDH branches are populated. The QDH exciton lives for $\tau_h$ nano-seconds and then both branches of the hybrid relaxes to the ground state emitting photons of the energy $\hbar \omega_{1S}\pm \Gamma_k$.
\par
Generalizing conventional double-sided Feynman diagrams \cite{MUKAMEL:1995} to include the non-radiative processes, the linear response from the QDH can be represented by the following diagram:
\begin{eqnarray}
\label{diagram}
\xymatrix{
\left|g\right\rangle & & \left\langle g\right| & \\
& \left\langle {1S \oplus F} \right\rangle \ar@{~>}[ul]^{\hbar \omega_{1S}-\Gamma_k} \ar@{~>}[ur]_{\hbar \omega_{1S}+\Gamma_k}& & \\
 \left|g\right\rangle \ar@{~>}[ur]_{\hbar \omega_{1S}} & & \left\langle {DCM2} \right| \ar[r] \ar[ul]^{\hbar \omega_{1S}} & \left|{CA}\right\rangle \ar@{<.>}[uu]^{\tau_h}\\
  & & \left\langle g\right| \ar@{~>}[u]^{E_{DCM2}} & \ar@{<.>}[u]^{\tau_s-\tau_h}}\\
  \nonumber
\chi^{\left(1\right)}_i \left(\omega,k\right)=\mu_{1S,0}^i \left({B^\dag_{1S,0}+B_{1S,0}}\right)+\mu_F^i \left({b^\dag_0+b_0}\right)=\\
= \frac{\left({\mu^i_F}\right)^2 \left({\hbar \omega - \hbar \omega_{1S} + i \hbar \gamma}\right) + \mu_F^i \mu^i_{1S} \Gamma_k}{\left({ \hbar \omega - \hbar \omega_{1S} +i \hbar \gamma}\right)^2 - \Gamma_k^2}+c.c.
\end{eqnarray}
On the diagram the wavy lines represent the incoming and outgoing photons; the straight lines stand for the non-radiative transitions. The diagram shows energy exchange between photon-exciton and exciton-exciton as well as the time separation between two pumping signals. Time increases from bottom to the top of the diagram as for the conventional Feynman diagram. The hybrid life time is denoted as $\tau_h= 1/\gamma$ and the hybridization between FE and WE is denoted as $\oplus$.
\par
In the derivation of the linear response $\chi^{\left(1\right)}_i \left(\omega,k\right)$ we used equation (\ref{EQ:3_1}) along with solutions of the Heisenberg equations of motion presented in the Appendix II (\ref{B0}). Formally the linear response can be written in terms of the hybrid Green's functions as:
\begin{eqnarray*}
\chi^{\left(1\right)}_i \left(\omega,k\right) = \sum\limits_{a,b = \left\{ {g,1S,F} \right\}} {\mu^i_{ab} \mu^i_{ba} I_{ab} \left( \omega  \right)} \\
I_{1S,g}  = I_{F,g}  = \frac{{\hbar \omega  - \hbar \omega_{1S}  + i\hbar \gamma }}{{\left( {\hbar \omega  - \hbar \omega_{1S}  + i\hbar \gamma } \right)^2  - \Gamma _k^2 }}\\
I_{1S,F} = \frac{\Gamma_k}{{\left( {\hbar \omega  - \hbar \omega_{1S}  + i\hbar \gamma } \right)^2  - \Gamma _k^2 }}\\
I_{ab}=I^\star_{ba}
\end{eqnarray*}
Here the dipole matrix elements in the corresponding basis (\ref{basis}) are given by:
\[
\left( {\begin{array}{*{20}c}
   0 & {\mu _{1S} } & {\mu _F } & 0 \\
   {\mu _{1S} } & 0 & {\sqrt {\mu _{1S} \mu _F } } & 0 \\
   {\mu _F } & {\sqrt {\mu _F \mu _{1S} } } & 0 & 0 \\
   0 & 0 & 0 & 0 \\
\end{array}} \right)
\]
Note that we neglected the non-resonant term associated with ground state dipole moment of the organic $\mu_g$.
\par
The SHG is due to second order response $E_j \bot E_i||z$ and given by the last term in the equation (\ref{EQ:3_1}) and the solutions of the equations of motion (\ref{B0},\ref{B1}). The first type of the SHG is formed when the branches of the hybrid interacts with the $\left|{2P}\right\rangle$ level excited by the prob signal. Using all the diagram conventions we adopted above, the diagram for this non linear process is given below:
\begin{eqnarray}
\nonumber
\xymatrix{
& & \left|g\right\rangle & \\
& \left|g\right\rangle &  & \left\langle g\right| & \\
\left|2P\right\rangle \ar@{<->}[rr]_{\mu_{1S,2P}} &  & \left\langle {1S \oplus F} \right\rangle \ar@{~>}[ul]|{\hbar \omega_{1S}-\Gamma_k} \ar@{~>}[ur]|{\hbar \omega_{1S}+\Gamma_k} \ar@{~>}[uu]|{\frac{\hbar \omega_{2P}}{2}} \ar@{~>}[dd]|{\frac{\hbar \omega_{2P}}{2}}& & \\
\left|g\right\rangle \ar@{~>}[u]_{\hbar \omega_{2P}} & \left|g\right\rangle \ar@{~>}[ur]_{\hbar \omega_{1S}} & & \left\langle {DCM2} \right| \ar[r] \ar[ul]^{\hbar \omega_{1S}} & \left|{CA}\right\rangle \ar@{<.>}[uu]^{\tau_h}\\
\ar@{<.>}[u]_{\tau_{2P}>\tau_s-\tau_h} &   & \left\langle g\right| & \left\langle g\right| \ar@{~>}[u]^{E_{DCM2}} & \ar@{<.>}[u]^{\tau_s-\tau_h}}\\
\chi^{\left(2\right)}_{ij} \left(2\omega;\omega,\omega\right) = \mu_{1S,2P} \left({B^\dag_{1S,0}B_{2P,1}+c.c.}\right)=\\
\frac{\mu^i_{2P} \mu^j_{1S,2P} \left({\mu^i_{1S}\left({\hbar \omega - \hbar \omega_{1S} + i \hbar \gamma}\right) + \mu^i_F \Gamma_k}\right)}{\left({2 \hbar \omega - \hbar \omega_{2P}}\right)\left({\left({\hbar \omega - \hbar \omega_{1S} + i \hbar \gamma}\right)^2-\Gamma^2_k}\right)}+c.c.
\end{eqnarray}
Here the prob signal comes \emph{after} the hybrid is formed: $\tau_{2P}>\tau_s-\tau_h$.
\par
Another second order non linear response can be formed if the prob signal is coming \emph{before} the hybridization $\tau_{2P}<\tau_s-\tau_h$.
It can be represented by the following diagram:
\begin{eqnarray*}
\nonumber
\xymatrix{
\left|g\right\rangle & \left|g\right\rangle &\left\langle g\right| & \\
 & & & \\
\left|g\right\rangle \ar@{~>}[r]^{\hbar \omega_{1S}} & \left\langle {1S \oplus F} \right\rangle \ar@{~>}[uul]|{\frac{{\hbar \omega_{1S}\pm \Gamma_k}}{2}} \ar@{~>}[uur]|{\frac{\hbar \omega_{1S}\mp \Gamma_k}{2}} \ar@{~>}[uu]|{\hbar \omega_{2P}} \ar@{<->}[d]|{\mu_{1S,2P}} &  &  \\
& \left|2P\right\rangle & \left\langle {DCM2} \right| \ar[ul] \ar[r] & \left|{CA}\right\rangle \ar@{<.>}[uuu]^{\tau_h} \ar@{<.>}[d]_{\tau_s-\tau_h} \\
\ar@{<.>}[u]_{\tau_{2P}<\tau_s-\tau_h} & \left\langle g\right|\ar@{~>}[u]_{\hbar \omega_{2P}} & \left|g\right\rangle \ar@{~>}[u]^{E_{DCM2}}&
}\\
\chi^{\left(2\right)}_{ij} \left(2\omega;\omega,\omega\right) = \mu_{1S,2P} \left({B^\dag_{1S,1}B_{2P,0}+c.c.}\right)=\\
= \frac{\mu^i_{2P} \mu^j_{1S,2P} \left({\mu^i_{1S}\left({2\hbar \omega - \hbar \omega_{1S} + i \hbar \gamma}\right) + \mu^i_F \Gamma_k}\right)}{\left({\hbar \omega - \hbar \omega_{2P}}\right)\left({\left({2\hbar \omega - \hbar \omega_{1S} + i \hbar \gamma}\right)^2-\Gamma^2_k}\right)}
\end{eqnarray*}
The Green's function representation of the SHG due to the second order response is given by the following expression:
\begin{eqnarray*}
\chi^{\left(2\right)}_{ij} \left(2\omega;\omega,\omega\right) =
\mu^j_{1S,2P} \sum\limits_{a = \left\{{g,1S,F,2P}\right\}} \mu^i_{a,1S} \mu^i_{2P,a} \times \\
\times \left[ {I_{a,1S} \left( \omega  \right)I_{a,2P} \left( {2\omega } \right) + I_{1S,a} \left( {2\omega } \right)I_{2P,a} \left( \omega  \right)} \right] \\
I_{2P,g}  = \frac{1}{{\hbar \omega  - \hbar \omega_{2P} }}
\end{eqnarray*}
The dipole matrix elements on the basis (\ref{basis}) are given by:
\[
\left( {\begin{array}{*{20}c}
   0 & {\mu _{1S} } & {\mu _F }  & {\mu_{2P}}\\
   {\mu _{1S} } & 0 & {\sqrt {\mu _{1S} \mu _F } } & {\mu_{1S,2P}} \\
   {\mu _F } & {\sqrt {\mu _F \mu _{1S} } } & 0 & 0 \\
   {\mu_{2P}} & {\mu_{1S,2P}} & 0 & 0
\end{array}} \right)
\]
According to the  last term in the equation (\ref{EQ:3_1}), the signal at $2 \hbar \omega = \hbar \omega_{1S}\pm \Gamma_k$ may generate the signal at $\hbar \omega = \hbar \omega_{1S}\pm \Gamma_k$:
\begin{eqnarray*}
\chi^{\left(3\right)}_{ij} \left(\omega;2\omega,-\omega\right)
= \frac{\mu^i_{2P} \left({\mu^j_{1S,2P}}\right)^2 \left({2\hbar \omega - \hbar \omega_{1S} + i \hbar \gamma}\right)}{\left({\hbar \omega - \hbar \omega_{2P}}\right)^2\left({\left({2\hbar \omega - \hbar \omega_{1S} + i \hbar \gamma}\right)^2-\Gamma^2_k}\right)}+c.c.
\end{eqnarray*}
This type of signal has been experimentally detected \cite{SHEN:1996} in bulk cuprous oxide ($\Gamma_k=0$) when the pumping signal was tuned to the wave length between $12285 \; \AA$ and $12195 \; \AA$. A strong SH signal was detected at $6096 \; \AA$ which has to be attributed not only to the narrow line-width of the quadrupole exciton but to the fact that $\mu_{1S,2P} \gg \mu_{1S}$ as well. From the last expression it follows that in this case no increment in the outgoing signal can be expected due to the hybridization effect.
\par
The third order nonlinearity is responsible as well for some small contribution to the SHG due to the non-zero ground state dipole moment of the DCM2 molecules \cite{KISHIDA:1994}. In the local electric field created by the polar CA molecules on the interface  $E_{loc}\left(0\right)$ ,the SH signal is due to the third order susceptibility $\chi^{\left(3\right)}_{ij} \left(2\omega;\omega,\omega,0\right)$. The exact expression in terms of the corresponding Green's functions is too lengthy to be listed here \cite{MUKAMEL:1995}, therefore we provide numerical calculations of the total SHG including the above correction in the next section.
%-----------------------------------------------------------------------------------------------------------------------------
\section{Results and discussion}
In order to make a numerical comparison of the hybrid and bulk SHG the life-time of the hybrid plays a major role. Considering the bi-stability effect in the hybrid \cite{ROSLYAK:2007} we assumed that the cuprous oxide has purity of $99.99 \%$ with the reported line-width of $\hbar \gamma_{1S} = 0.1 \; meV$ (pico-second lifetime) \cite{SHEN:1996}. Therefore the hybrid life-time is dominated by its inorganic part $\hbar \gamma \approx \hbar \gamma_{1S}$. To compensate for such big line-width we also assumed that the DCM2 is presented as a thin film embedded into PS host close to the interface with the cuprous oxide. For the non-linear absorption experiment this assumption can be justified as it makes the absorption length of the hybrid equal to the narrow region around the interface, of the size of the hybrid itself. But there is a drawback in that model due to possible aggregation of the DCM2.
\par
Hence in this article we adopted the picture of disordered organic and higher purity of the inorganic crystal. This will bring the line-width and the coupling parameter to the same order. For pure cuprous oxide crystal the life-time of the quadrupole $1S$ exciton is reported to be $\tau_{1S} = 1.7 \ldots 3.0 \; ns $ ($ \hbar \gamma = 1 \ldots 0.5 \; \mu eV$) \cite{DASBACH:2004,FROHLICH:2005,ELLIOT:1961}. Such crystals and thin films are widely used in searching for BEC of excitons.
\par
In this case the life-time of the $1S$ quadrupole exciton  is mainly determined by the ortho-para exciton conversion. The life-time of the organic part of the hybrid is determined by the time the excited DCM2 molecule reaches an equilibrium with the bath of polar CA molecules. The life-time for the given concentration of the CA is reported to be $3.3 \; ns$ \cite{MADIGAN:2003,MADIGAN:2004}. Because these processes are of the same order, the effective life-time of the hybrid is a non-trivial combination of the effects described above and will be reported elsewhere. Here we assume the simplest case of non-coherent life-time of the hybrid $\hbar \gamma = 0.29 \; \mu eV$ \cite{ROSLYAK:2006}.
\par
\begin{figure}[htbp]
\centering
\includegraphics[width=8.6cm]{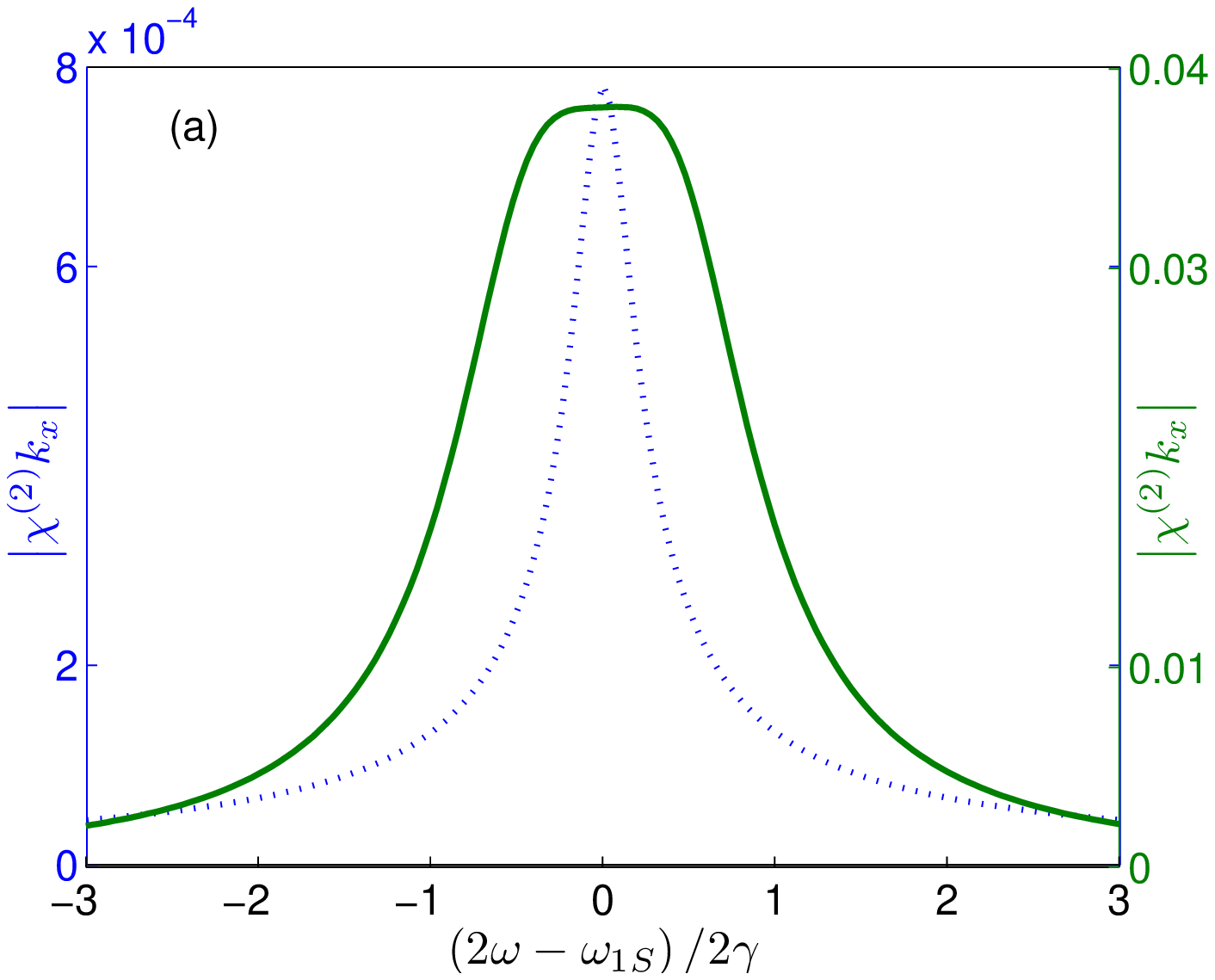}
\includegraphics[width=8.6cm]{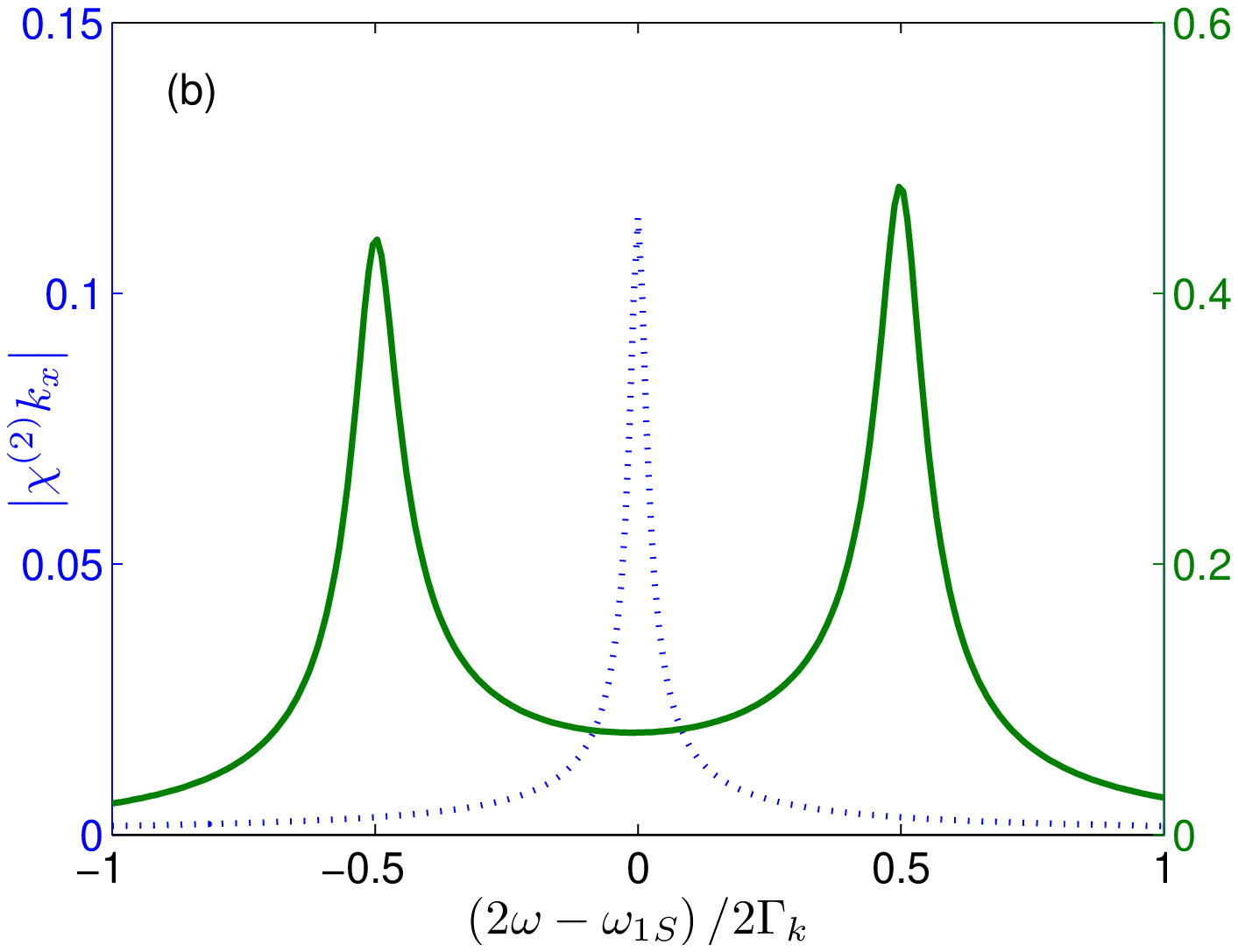}
	\caption{(Color on-line) Relative value of the nonlinear susceptibility in case of bulk cuprous oxide (dotted curves) and the quadrupole-dipole hybrid (solid curves). The density of the disordered DCM2 is taken $\rho_{DCM2}=0.005 \%$ while the CA density is $\rho_{CA}=22 \%$. The Fig.2a represents moderate coupling $\Gamma_k = \hbar \gamma_{1S} =0.29 \; \mu eV$ and Fig.2b corresponds to strong coupling regime $\Gamma_k = 3.5 \; \mu eV$. In the last case the enhancement is evident and indicated by the different scales for the bare cuprous oxide (left) and hybrid (right) SHG}
	\label{FIG:2}
\end{figure}

The intensity of the second-harmonic is proportional to $\left|{\chi^{\left({2}\right)}k_x}\right|^2$ (See for example \cite{HAUEISEN:1973}). Therefore an important measurable quantity is a relative value of nonlinear susceptibility $\left|{\chi^{\left({2}\right)}k_x}\right|$ presented in Fig.\ref{FIG:2}.
% FIGURE SUSCEPTIBILITY BEGIN
% FIGURE SUSCEPTIBILITY END
\par
The SHG signal is split according to the response from the lower and upper branch of the hybrid. Asymmetry between this two branches is a result of quantum effects and not present in the classical anharmonic oscillator picture. We also included the corrections due to interface effect in the organic in our numerical simulation.
\par
For the sake of simplicity let us consider two distinct cases. First, the pump laser is perpendicular to the interface. The states up to $ka = k_0a$ are populated thermally. No hybridization occurs and it is equivalent to the bulk case SHG (See Fig.\ref{FIG:2} dotted curve). The maximum power generated by the second-harmonic is proportional to the square of the following expression:
\begin{equation}
\label{EQ:4_1}
\left|{\chi^{\left({2}\right)}_{ij,max}\left({2\hbar \omega = \hbar \omega_{1S}}\right)}\right| = \frac{\mu_{2P} \mu_{1S,2P}}{\hbar \omega - \hbar \omega_{2P}} \frac{\mu_{1S,k} k_x}{\hbar \gamma_{1S}}
\end{equation}
The small relative value of the SHG is due to the narrowness of the cuprous oxide quantum well.
\par
Second, the pump laser incidence angle is reduced to acquire the wave vector $k_0a \ll ka$.
The maximum power generated by the second-harmonic is proportional to the square of the following expression:
\begin{equation}
\label{EQ:4_2}
\left|{\chi^{\left({2}\right)}_{ij,max}\left({2\hbar\omega = \hbar \omega_{1S} \pm \Gamma_k}\right)}\right| = \frac{\mu_{2P} \mu_{1S,2P}}{\hbar \omega - \hbar \omega_{2P}} \frac{\mu_{F} k_x}{\alpha_k \hbar \gamma}
\end{equation}
Here the incidence angle dependent coefficient $\alpha_k = \sqrt{5}$ (See Fig.\ref{FIG:2}a) for $ka = 0.13$ ($\Gamma_k = \hbar \gamma =0.29 \; \mu eV$) and $\alpha_k = 2$ for maximum value of the coupling $\Gamma_k = 3.5 \; \mu eV$ at $ka=1.57$ (See Fig.\ref{FIG:2}b).
Finally, comparing the last expression (\ref{EQ:4_2}) to the bulk cuprous oxide (\ref{EQ:4_1}) the second order response of the hybrid is amplified by the factor:
\[\left({\frac{\mu_F}{\mu_{1S}} \frac{\hbar \gamma_{1S}}{\alpha \hbar \gamma}}\right)^2\]
Therefore the amplification can be adjusted by manipulating the organic composition (DCM2 and CA densities) or changing the pump laser incidence angle.
\par
Finally, we would like to note that there is another merit in using the hybrid structure for the SHG. Namely the fact that the optical pumping can be replaced by an electrical pumping. For this the hybrid sample has to be placed between Alq3 and a-NPD \cite{MADIGAN:2004} semiconductor plates. The bond structure and offset of these materials provide electrons and holes to form the hybrid exciton on the interface. Although, in this case one can expect the SHG only from the lower branch of the hybrid as the excitons are accumulated in the minimum of the hybrid dispersion \cite{ROSLYAK:2006}.
%----------------------------------------------------------------------------------------------------------------------------
\section{Conclusion}
In this paper we addressed possibility to enhance SHG signal $\chi^{(2)}$ generic to a cuprous oxide bulk crystal as the lowest excitation in this material has quadrupole origin. To demonstrate the concept we proposed to consider a pump-prob experiment performed on cuprous oxide sandwiched between the organic composite. An intense pump signal excites one part of the organic known as DCM2 (FE). Non-resonant (F{\"o}rster) energy transfer in the organic layer ("solid state solvation" effect) provides dynamical red shift of the FE. When the FE energy is close enough to the quadrupole allowed 1S exciton in the adjacent cuprous oxide the quadrupole-dipole hybridization occurs due to the FE induced gradient of the electric field penetrating into the inorganic layer. The prob signal is designed to reveal the SHG signal.
\par
The resonant enhancement of the $\chi^{(2)}$ occurs for the hybrid exciton shares properties of both quadrupole WE (long radiative lifetime) and organic FE (big oscillator strength). I't quadrupole part allows $\chi^{(2)}$ to be non-vanishing. While it's FE part provides the enhancement of the SHG signal compared to the bare cuprous oxide crystal due to more efficient absorption of the pump signal by means of large oscillator strength of the hybrid.
\par
However, as we demonstrated in the classical coupled oscillator model framework,the enhancement is determined not only by the big ratio of the corresponding organic/inorganic oscillator strengths but somehow quenched by the small coupling parameter and low DCM2 density. By varying those parameters one the hybrid SHG signal may be enhanced by orders of magnitude compared with the generic (cuprous oxide) one.
\par
To reveal the enhancement dependence on such an important parameter of the hybrid as the hybridization time $\tau_h$ we proposed more sophisticated quantum theory. It suggests that there is substantial difference in the hybrid SHG signal provided one probes the system before or after the hybridization occurred. In the first case the SH is generated at $\omega_{2P}/2$ frequency. Hence it is vastly suppressed by the short life time of the dipole-allowed 2P WE in cuprous oxide. Nevertheless, if one probes the system after the hybridization had happened, the SH is generated at $\omega_{1S}/2$  and is heightened by the big oscillator strength of the hybrid and it's small damping coefficient.
%----------------------------------------------------------------------------------------------------------------------------
\section{Acknowledgments}
We would like to thank Ms. Upali Aparajita for helpful discussion and comments. The project was supported in part by PCS-CUNY.
%----------------------------------------------------------------------------------------------------------------------------

%----------------------------------------------------------------------------------------------------------------------------
\appendix
%----------------------------------------------------------------------------------------------------------------------------
\section{}
Explicit expression for the quadrupole-dipole coupling is given below:
\begin{equation}
\label{EQ:1_1}
\Gamma_{k}=\frac{8\sqrt {2\pi } }{\left(
{\varepsilon + \tilde {\varepsilon }} \right)L_w
}\frac{ke^{-k{z}'}\texttt{sinh}\left( {\frac{L_w k}{2}} \right)}{\left( {1+\left(
{\frac{kL_w }{2\pi }} \right)^2} \right)}\frac{Q_{xz} \mu_z^F }{a^F_B
a^W_B L_w }
\end{equation}
Here $a^F_B$, $a^W_B$ are the Bohr radius of the FE and WE exciton; $\tilde {\varepsilon }$ and $\varepsilon$ are the corresponding dielectric constants, $z'$ is the distance to the DCM2 layer, $L_w$ is the quantum well width. The quadrupole transition matrix element $Q_{xz}$ may be estimated from the corresponding oscillator strength per unit cell through the following identity \cite{MOSKALENKO:2002} and depends on polarization of the pumping laser field:
\begin{eqnarray}
\label{EQ:1_0}
f_{xz,k_0} =\frac{4\pi m E_g }{3e^2\hbar ^2}\left( {\frac{a^W_B
}{a}} \right)^3\left({{\rm {\bf z}} \cdot {\rm {\bf k}}_{0,x} \cdot Q_{x,z}}\right)^2\\
\nonumber
f_{xz,{\rm {\bf k}}_0 || [1,1,0] } =3.9\times 10^{-9}\\
\nonumber
f_{xz,{\rm {\bf k}}_0 || [1,1,1]} =\frac{1}{3} 3.9\times 10^{-9}
\end{eqnarray}
Here the energy gap of cuprous oxide is denoted as $E_g = 2.173 \; eV$; $k_0=2.62\times 10^5 \; cm^{-1}$ is the resonant wave vector; $a$ is the unit cell size; the unit vector in the pumping field polarization is ${\rm {\bf z}}$.
%-----------------------------------------------------------------------------------------------------------------------------
\clearpage
\section{}
The non-zero commutator relations for the organic and inorganic parts of the hybrid yield \cite{MUKAMEL:1995}:
\begin{eqnarray}
\label{algebra}
\fl \left[ {B_{1S}^\dag  ,B_{1S} } \right] =  - 1 + B_{2P}^\dag  B_{2P}  + b^\dag  b; & \left[ {B_{1S}^\dag  ,B_{2P} } \right] = B_{1S}^\dag  B_{2P}; & \left[ {B_{2P}^\dag  ,B_{1S} } \right] = B_{2P}^\dag  B_{1S} \\
 \nonumber
\fl \left[ {B_{2P}^\dag  ,B_{2P} } \right] =  - 1 + B_{1S}^\dag  B_{1S}  + b^\dag  b; & \left[ {b^\dag  ,B_{1S} } \right] = b^\dag  B_{1S}; & \left[ {b^\dag  ,B_{2P} } \right] = b^\dag  B_{2P} \\
 \nonumber
\fl \left[ {b^\dag  ,b} \right] =  - 1 + B_{2P}^\dag  B_{2P}  + B_{1S}^\dag  B_{1S};  & \left[ {B_{1S}^\dag  ,b} \right] = B_{1S}^\dag  b; & \left[ {B_{2P}^\dag  ,b} \right] = B_{2P}^\dag  b
 \end{eqnarray}
In the TDHF approximate factorization for the averages, the corresponding Heisenberg equations up to the second order in the creation and annihilation operators are:
\begin{eqnarray*}
\fl i \hbar \frac{d B^\dag_{1S}}{d t} = \hbar \omega_{1S} B^\dag_{1S} + \Gamma_k b^\dag - \mu^F E_i B^\dag_{1S} b + \mu_{1S,k} E_i \left({1 - B^\dag_{2P} B_{2P} - b^\dag b}\right) - \\
\mu_{2P} E_i B^\dag_{1S} B_{2P} + \mu_{1S,2P} E_j B^\dag_{2P} \\
\fl i \hbar \frac{d B_{1S} }{d t} = -\hbar \omega_{1S} B_{1S} - \Gamma_k b + \mu^F E^\star_i b^\dag B_{1S} - \mu_{1S,k} E^\star_i \left({1 - B^\dag_{2P} B_{2P} - b^\dag b}\right) + \\
\mu_{2P} E^\star_i B^\dag_{2P} B_{1S} - \mu_{1S,2P} E^\star_j B_{2P} \\
\fl i \hbar \frac{d B^\dag_{2P}}{d t} = \hbar \omega_{2P} B^\dag_{2P} - \mu^F E_i B^\dag_{2P} b - \mu_{1S,k} E_i B^\dag_{2P} B_{1S} + \mu_{2P} E_i \left({1 - B^\dag_{1S} B_{1S} - b^\dag b}\right) + \\
+\mu_{1S,2P} E_j B^\dag _{1S} \\
\fl i \hbar \frac{d B_{2P}}{d t}  = -\hbar \omega_{2P} B_{2P} + \mu^F E^\star_i b^\dag B_{2P} + \mu_{1S,k} E^\star_i B^\dag_{1S}B_{2P} - \mu_{2P} E^\star_i \left({1 - B^\dag_{1S} B_{1S} - b^\dag b}\right) - \\
-\mu_{1S,2P} E^\star_j B_{1S} \\
\fl i \hbar \frac{d b^\dag}{d t} = E^F b^\dag + \Gamma_{k} B^\dag_{1S} + \mu^F E_i \left({1 - B^\dag_{1S} B_{1S} - B^\dag_{2P} B_{2P}}\right) - \mu_{1S,k} E_i b^\dag B_{1S} - \mu_{2P} E_i b^\dag B_{2P}\\
\fl i \hbar \frac{d b}{d t}  = -E^F b - \Gamma_{k} B_{1S} - \mu^F E^\star_i \left({1 - B^\dag_{1S} B_{1S} - B^\dag_{2P} B_{2P}}\right) + \mu_{1S,k} E^\star_i B^\dag_{1S} b + \mu_{2P} E^\star_i B^\dag_{2P} b
\end{eqnarray*}
Here we omitted the average brackets for the shorter notation. In the exact resonance between FE and WE excitons $\hbar \omega_{1S}=\hbar \omega_F$ the linear approximation is straightforward. The creation operators are proportional to $\propto e^{i \omega t}$ and the system above is reduced to:
%\begin{widetext}
\begin{eqnarray*}
\hbar \omega B^\dag_{1S,0} = \left({\hbar \omega_{1S} - i \hbar \gamma }\right) B^\dag_{1S,0} + \Gamma_k b^\dag_0 + \mu_{1S,k} E_i -\hbar \omega B_{1S,0}  \\
= -\left({\hbar \omega_{1S} + i \hbar \gamma }\right) B_{1S,0} - \Gamma_k b_0 - \mu_{1S,k} E^\star_i \\
\hbar \omega B^\dag_{2P,0} = \hbar \omega_{2P} B^\dag_{2P,0} + \mu_{2P} E_i -\hbar \omega B_{2P,0}  \\
= -\hbar \omega_{2P} B_{2P,0} - \mu_{2P} E^\star_i  \\
\hbar \omega b^\dag_0 = \left({\hbar \omega_{1S} - i \hbar \gamma }\right) b^\dag_0 + \Gamma_{k} B^\dag_{1S,0} + \mu^F E_i-\hbar \omega b_0  \\
= -\left({\hbar \omega_{1S} + i \hbar \gamma }\right) b_0 - \Gamma_{k} B_{1S,0} - \mu^F E^\star_i
\end{eqnarray*}
\clearpage
The system above has a solution:
\begin{eqnarray}
\label{B0}
B^\dag_{2P,0} = \frac{\mu_{2P} E_i}{\hbar \omega - \hbar \omega_{2P}}\\
\nonumber
B^\dag_{1S,0} = \frac{\mu_{1S} E_i \left({\hbar \omega - \hbar \omega_{1S} + i \hbar \gamma}\right) + \mu^F \Gamma_k E_i}{\left({\hbar \omega - \hbar \omega_{1S} +i \hbar \gamma}\right)^2 - \Gamma_k^2} \\
\nonumber
b^\dag_0 = \frac{\mu^F E_i \left({\hbar \omega - \hbar \omega_{1S} + i \hbar \gamma}\right) + \mu_{1S} \Gamma_k E_i}{\left({\hbar \omega - \hbar \omega_{1S} +i \hbar \gamma}\right)^2 - \Gamma_k^2}
\end{eqnarray}
The SHG is due to response to induced polarization and is proportional to $\propto e^{i 2 \omega t}$:
\begin{eqnarray*}
2 \hbar \omega B^\dag_{1S,1} = \left({\hbar \omega_{1S} - i \hbar \gamma }\right) B^\dag_{1S,1} + \Gamma_k b^\dag_1 + \mu_{1S,2P} E_j B^\dag_{2P,0}  \\
2 \hbar \omega B_{1S,1} = \left({\hbar \omega_{1S} + i \hbar \gamma }\right) B_{1S,1} + \Gamma_k b_1 + \mu_{1S,2P} E^\star_j B^\dag_{2P,0} \\
2 \hbar \omega B^\dag_{2P,1} = \hbar \omega_{2P} B^\dag_{2P,1} + \mu_{1S,2P} E_j B^\dag_{1S,0}  \\
2 \hbar \omega B_{2P,1} = \hbar \omega_{2P} B_{2P,1} + \mu_{1S,2P} E^\star_j B^\dag_{1S,0}  \\
2 \hbar \omega b^\dag_1 = \left({\hbar \omega_{1S} - i \hbar \gamma }\right) b^\dag_1 + \Gamma_{k} B^\dag_{1S,1}  \\
2 \hbar \omega b_1 = \left({\hbar \omega_{1S} + i \hbar \gamma }\right) b_1 + \Gamma_{k} B_{1S,1}
\end{eqnarray*}
The system has a solution:
\begin{eqnarray}
\label{B1}
B^\dag_{2P,1} = \frac{\mu_{1S,2P} E_j B^\dag_{1S,0}}{2 \hbar \omega - \hbar \omega_{2P}}\\
\nonumber
B^\dag_{1S,1} = \frac{\mu_{1S,2P} E_j \left({2 \hbar \omega - \hbar \omega_{1S} + i \hbar \gamma}\right) B^\dag_{2P,0}}{\left({2 \hbar \omega - \hbar \omega_{1S} +i \hbar \gamma}\right)^2 - \Gamma_k^2} \\
\nonumber
b^\dag_{1} = \frac{\mu_{1S,2P} E_j \Gamma_k B^\dag_{2P,0}}{\left({2 \hbar \omega - \hbar \omega_{1S} +i \hbar \gamma}\right)^2 - \Gamma_k^2}
\end{eqnarray}
This solutions are implemented in the main text to calculate the linear and nonlinear responses of the hybrid.
%-----------------------------------------------------------------------------------------------------------------------------
\end{document}